\documentclass[onecolumn,a4paper,aps,prd,floats,floatfix,amsmath,amssymb,amsfonts,nofootinbib,longbibliography,superscriptaddress,10pt]{revtex4}
\usepackage[utf8]{inputenc}
\usepackage[T1]{fontenc}
\usepackage{graphicx}
\usepackage{xcolor}
\usepackage{natbib}
\usepackage[colorlinks]{hyperref}

\begin{document}


\title{A new framework for cosmological expansion in a reformulated Newtonian-Like gravity with variable G}

\author{Felipe S. Escórcio}
\affiliation{%
Departamento de F\'isica, Universidade Federal de Ouro Preto (UFOP), Campus Universit\'ario Morro do Cruzeiro, 35.400-000, Ouro Preto, Brazil}%

\author{J\'ulio C. Fabris}
\email{julio.fabris@cosmo-ufes.org}%
\affiliation{%
N\'ucleo Cosmo-ufes \& Departamento de F\'isica,  Universidade Federal do Esp\'irito Santo (UFES)\\
Av. Fernando Ferrari, 540, CEP 29.075-910, Vit\'oria, ES, Brazil.}%

\affiliation{National Research Nuclear University MEPhI (Moscow Engineering Physics Institute),
115409, Kashirskoe Shosse 31, Moscow, Russia}

\author{Júnior D. Toniato}%
\email{junior.toniato@ufes.br}
\affiliation{%
N\'ucleo Cosmo-ufes \& Departamento de Química e Física, Universidade Federal do Espírito Santo - Campus Alegre, ES, 29500-000, Brazil}%

\author{Hermano Velten}%
\email{hermano.velten@ufop.edu.br}
\affiliation{%
Departamento de F\'isica, Universidade Federal de Ouro Preto (UFOP), Campus Universit\'ario Morro do Cruzeiro, 35.400-000, Ouro Preto, Brazil}%
\date{\today}

\begin{abstract}

A Newtonian-like theory inspired by the Brans-Dicke gravitational Lagrangian has been recently proposed in Ref. \cite{jjtv}. We propose here a new variant of this theory such that the usual Newtonian second law is preserved. The cosmological solutions are analysed and accelerated background expansion can be obtained even in a pure matter dominated universe. This happens due to the dynamic character of the effective gravitational coupling which is sourced by a time evolving scalar field . We also analyse the matter density perturbations and find they exhibit an enhanced growth in comparison with the usual Newtonian like behavior in Einstein-de Sitter model. 
\end{abstract}

\maketitle

\section{Introduction}

The understanding of the evolution of the universe is one of the central pillars of cosmology, and with the consolidation of Einstein's General Theory of Relativity, a robust theoretical foundation began to emerge. Einstein, in developing his field equations, initially postulated a static universe, formalized by the ``cosmological constant'' as a necessity to balance gravitational attraction preventing the collapse of the universe. In 1922, Alexander Friedmann presented solutions to Einstein's equations that allowed for a dynamic universe, introducing the possibility of a cosmos that could either expand or contract. Empirical confirmation of this solution came with Edwin Hubble's observations in 1929, which demonstrated that galaxies were moving away from each other, corroborating the previous inference from Georges Lemaître, and firmly establishing the expansion of the universe.

Since then, SNe type Ia observations have revealed a new aspect of the expansion: It is also accelerated. This is attributed to the presence of dark energy, which appears as an unknown form of energy that, according to the cosmological concordance model, constitutes approximately $68\%$ of the total energy density of the universe \cite{Planck2018}, and which remains poorly understood to this day. Moreover, even being the universe described on cosmological scales by a homogeneous and isotropic distribution, the formation of large-scale structures, such as galaxy clusters and superclusters, is understood as the result of the gravitational amplification of the primordial small density perturbations.

The density fluctuations, which originated during the inflationary era, have been detected and precisely mapped through anisotropies in the cosmic microwave background radiation. However, in order to obtain a successful description of the large scale structure a dark matter component is also demanded. It constitutes approximately 27\% of the total energy density. Its composition and properties remain unknown, being inferred indirectly through gravitational effects. Dark energy also presents a significant challenge to fundamental physics, as its nature is not compatible with the expected values for Einstein's cosmological constant. More specifically, there is a significant discrepancy between the predicted values for this constant within the framework of quantum field theory and the observational values obtained from cosmological measurements \cite{Martin}. This divergence highlights a critical gap in the unification of these two fundamental theories.

The absence of a convincing explanation for these and other cosmological issues has motivated the investigation of new approaches for the gravitational sector \cite{CANTATA:2021asi}. The exploration of alternative theories to General Relativity, such as Brans-Dicke theory \cite{Brans:1961sx} and other modified gravity proposals, could pave the way for discoveries that clarify the true nature of the dark sector. These alternative theories seek not only to describe gravitational behavior in distinct regimes but also to offer testable predictions that can be compared with cosmological and astrophysical observations. In this regard, the Brans-Dicke theory was one of the first attempts to suggest the possibility that the gravitational coupling $G$ may vary in certain regimes, with the specific form of this variation being determined within the model itself. The mechanism behind this variation is encoded in the existence of a new scalar field which, together with the metric field, also mediates the gravitation interaction. The Horndeski's theories represent a larger class of scalar-tensor theories and present a richer phenomenology in explaining both the background expansion and the structure formation process \cite{Horndeski:1974wa}. 

Whether the fundamental constants of physics truly remain constant has been a topic of interest. Among these, four constants hold particular significance: $h$, which governs quantum phenomena; $c$, the speed of light, key to relativistic effects; $G$, representing gravitational interaction; and $k_B$, the Boltzmann constant, central to thermodynamics. Of these, $G$ was the first to be discovered, yet it remains the least precisely known, with an accuracy only up to the order of $10^{-4}$ \cite{10.1093/nsr/nwaa165,microscope}.

Although there are several covariant relativistic theories that predict a variable gravitational coupling, the formulation of a Newtonian theory that incorporates a dynamic $G$ faces considerable challenges due to the difficulty in integrating a variable gravitational coupling into a Newtonian framework that, by definition, assumes a fixed and universal $G$. The dynamic nature of $G$ would require a reformulation of Newtonian theory, implying the need for a consistent and precise description in this context, with the existence of a more fundamental description that justifies its variability. 
A non-relativistic approach for gravitational systems is justified since many problems in astrophysics and cosmology are well described by the Newtonian approach.  Early attempts to incorporate a varying $G$ in a Newtonian framework were relatively straightforward, substituting the constant $G$ in the Poisson equation by a time-dependent gravitational coupling function $G \equiv G(t)$ \cite{Dirac:1937ti,Dirac:1938mt, Duval:1990hj}. In all such variable $G$ formulations so far the $G(t)$ function is imposed {\it ad hoc}, with no dynamic equation to determine its temporal evolution.

In recent works \cite{jjtv,Fabris:2021qkp,Escorcio:2023gnu}, a new Newtonian theory with a variable gravitational coupling has been proposed. In this approach, the gravitational coupling is described in terms of a new field $\sigma$ which depends on both time and position, $\sigma \equiv \sigma(t,r)$. Its dynamics is determined from a Lagrangian function along with the gravitational potential $\psi$. This formulation allows for the derivation of an expression for the advance of the orbital pericenter and the consequent constraint on the model free parameter $\omega$ (similar to the Brans-Dicke parameter). Additionally, the variability of the gravitational coupling naturally emerges from the proposed Lagrangian, demonstrating that the theory successfully reproduces the advance of Mercury's perihelion without significantly impacting the Roche limit when compared to the results predicted by Newtonian celestial mechanics \cite{Escorcio:2023gnu}. Apart from this, it is worth mentioning that numerical simulations of the cosmological large scale structure are very useful tools to learn how galaxies and clusters have evolved along universe's lifetime. Such simulations are based mainly on Newtonian gravity and non-relativistic hydrodynamics. At the same time, modified gravity theories are formulated based on a covariant description of geometrical properties of space-time.

In this work, we develop a new prototype of this variable $G$ Newtonian-like gravitational theory and apply it to the scenario of the universe's expansion, with additional focus on perturbative calculations aimed at understanding the evolution of large-scale structure formation. Our main objective is to verify whether it is possible to adjust the Lagrangian proposed in previous works to make it more suitable to treat cosmological perturbations.

To this end, we review the general theoretical framework at the beginning of the next section, establishing the conceptual and mathematical foundations of the theory, as well as the calculations performed in previous works. We then perform the necessary corrections in the initial model and investigate the behavior of these corrections in relation to the predicted cosmological solutions. Subsequently, we present the calculation of perturbations and discuss the different possible expansion results justified by the variability of the gravitational constant, according to the values assigned to the free parameter $\omega$.

\section{Newtonian theory with variable $G$: first formulation}\label{sec:fisrt}

In Ref. \cite{jjtv} a formulation of a Lagrangian based Newtonian theory with variable $G$ has been proposed. Subsequently, in Ref. \cite{Fabris:2021qkp}, it was estimated the impact of such model to stellar structure. More recently, a more precise study revealed all the nuances of this modified Newtonian gravity, establishing constraints from periastron advance observational data, while discussing more subtle subjects as the possible breaking of the equivalence between inertial and gravitational masses \cite{Escorcio:2023gnu}.
The Lagrangian proposed in Ref. \cite{jjtv} is written as follows, 
\begin{eqnarray}
\label{LagrangianG}
{\cal L} = -\frac{\nabla\psi\cdot\nabla\psi}{8\pi G_0} + \frac{\omega}{8\pi G_0}\biggr(\psi\frac{\dot\sigma^2}{\sigma^2} - c^4\nabla\sigma\cdot\nabla\sigma \biggl) - \,\rho \sigma \psi,
\end{eqnarray}
with $\psi$ and $\sigma$ being scalar functions, $\rho$ representing the matter density, $\omega$ is a dimensionless constant and $G_0$ a constant with same dimensions of Newton's gravitational constant $G_N$. Although the constant $c$ indicates light velocity in vacuum, it appears as a convenient way to correct units while leaving $\sigma$ dimensionless. It does not necessarily evoke the usual notion of signal speed limit as imposed by special relativity, and it can be interpreted as a constant with velocity dimensions constructed from electromagnetic quantities, the permittivity and permeability in vacuum.

From the above Lagrangian, by applying Euler-Lagrange equations of motion, one can derive the
scalar fields dynamics,
\begin{eqnarray}
\label{eqno1}
\nabla^2\psi + \frac{\omega}{2}\left(\frac{\dot\sigma}{\sigma}\right)^2  &=& 4\pi G_0 \sigma \rho,\\[1ex]
\label{eqno2}
\nabla^2\sigma - \frac{1}{c^4\sigma}\frac{d}{dt}\left(\frac{\psi\dot\sigma}{\sigma}\right) &=& \dfrac{4\pi G_0\psi \rho}{c^4\omega}.
\end{eqnarray}
Due to the direct coupling between $\rho$ and the fields $\psi$ and $\sigma$, the effective gravitational potential able to influence matter dynamics is given by the product of both fields, as shown by the equations of motion satisfied by an auto-gravitating fluid,
\begin{eqnarray}
\label{eqno3}
\frac{\partial\rho}{\partial t} + \nabla\cdot(\rho\vec v) &=& 0,\\
\label{eqno4}
\frac{\partial \vec v}{\partial t} + \vec v\cdot\vec\nabla v &=& - \frac{\vec\nabla p}{\rho} - \vec\nabla(\sigma\psi).
\end{eqnarray}
It is worth to note that such theory recovers Newtonian gravity in the limit where both $\sigma\equiv \sigma_0$ is constant and $\omega$ goes to infinity, with $G_0\sigma_0$ playing the role of the gravitational constant. In this limit, by assigning $\sigma_0=1$, then $G_0$ refers to the exact value of $G_N$.

In order to describe the cosmological context, the velocity field associated to the homogeneous and isotropic expansion of the universe is written as
\begin{eqnarray}\label{vHr}
\vec v = \frac{\dot a}{a}\vec r,
\end{eqnarray}
where $\vec r = a(t) \vec r_c$ is the physical distance related to the comoving distance $\vec r_c$ with $a\equiv a(t)$ being a function of time which is identified subsequently with the cosmological scale factor. This is the well known Hubble-Lemaître law and the expansion rate is defined as
\begin{equation}
    H\equiv \frac{\dot{a}}{a}.
\end{equation}

Moreover, given the dynamical cosmological background, both the matter density $\rho$, the pressure $p$ and the field $\sigma$ must also be functions of the time coordinate only. Given this velocity law, the conservation equation (\ref{eqno3}) can be integrated, leading to,
\begin{eqnarray}\label{density}
\rho = \frac{\rho_0}{a^3}.
\end{eqnarray}

The equations (\ref{eqno1}-\ref{eqno4}) do not admit power law solutions as in the traditional Newtonian case. In fact, combining (\ref{eqno1}) and (\ref{eqno2}), it is possible to verify that the potential $\psi$, under the hypothesis of a power law behavior, must scale as $t^{-2}$. This will not be consistent with (\ref{eqno4}) unless $\sigma$ is constant and $\omega \rightarrow \infty$ [because of eq. (\ref{eqno2})], recovering the original Newtonian equations with a constant gravitational coupling.

Before starting to discuss a new proposal for modifying Newtonian gravity, let us develop a little bit further the cosmological equations. With a redefinition of the $\psi$ field,
\begin{eqnarray}
\psi = g(t)\frac{r^2}{6},
\end{eqnarray}
we can write equations (\ref{eqno1}-\ref{eqno4}) as follows,
\begin{eqnarray}
g + \frac{\omega}{2}\frac{\dot\sigma^2}{\sigma^2} &=& 4\pi G_0\rho_0\frac{\sigma}{a^3},\\[1ex]
 \frac{\ddot\sigma}{\sigma}+\biggr(2H+\frac{\dot g}{g} \biggl)\frac{\dot\sigma}{\sigma}  - \frac{\dot\sigma^2}{\sigma^2} &=& - 
\frac{4\pi G_0\rho_0}{\omega}\frac{\sigma}{a^3},\\[1ex]
\dot{H}+H^2 &=& - \sigma\frac{g}{3},
\end{eqnarray}
with
\begin{eqnarray}\label{g(t)}
g(t) = 4\pi G_0\sigma\rho - \frac{\omega}{2}\frac{\dot\sigma^2}{\sigma^2}.
\end{eqnarray}

In the next section we seek for cosmological solutions of the above system of equations.

\section{Newtonian theory with variable $G$ revisited}

Let us discuss now a new, different formulation of the Newtonian theory with variable $G$. The idea is to have no direct coupling of the field $\sigma$ with matter, at the price of imposing a coupling with the gradient of the gravitational potential $\psi$. 
A concrete possibility is given by the new Lagrangian,
\begin{eqnarray}
{\cal L} = - \frac{\nabla\psi\cdot\nabla\psi}{8\pi G_0\sigma} + \frac{\omega}{8\pi G_0}\biggr(\psi\frac{\dot\sigma^2}{\sigma^3} - c^4\frac{\nabla\sigma\cdot\nabla\sigma}{\sigma}\biggl) - \rho\psi. \label{new-action}
\end{eqnarray}
In this Lagrangian $G_0$ is a reference constant value for the gravitational coupling, as the value of $G$ today. Now, the resulting equations for the self-gravitating system are given by,
\begin{eqnarray} \label{eqsnewpsi}
\nabla^2 \psi - \frac{\nabla\psi\cdot\nabla\sigma}{\sigma} + \frac{\omega}{2}\frac{\dot\sigma^2}{\sigma^2} &=& 4\pi G_0\sigma\rho,\\
\label{eqsnewsigma}
\ddot\sigma - \frac{3}{2}\frac{\dot\sigma^2}{\sigma} + \frac{\dot\psi}{\psi}\dot\sigma + c^4 \frac{\sigma^2}{\psi} \biggr(- \nabla^2\sigma
+ \frac{\nabla\sigma\cdot\nabla\sigma}{2\sigma}\biggl) - \frac{\sigma}{2\omega}\frac{\nabla\psi\cdot\nabla\psi}{\psi} &=& 0,\\
\frac{\partial\rho}{\partial t} + \nabla\cdot(\rho\vec v) &=& 0,\\
\label{eqeuler} \frac{\partial \vec v}{\partial t} + \vec v\cdot\nabla \vec v &=& - \frac{\nabla p}{\rho} - \nabla\psi.
\end{eqnarray}


Lagrangian \eqref{new-action} and the consequent field equations allow us to better understand the role played by each one of the fields in this new formulation. Through equation \eqref{eqeuler} it is seen that $\psi$ plays the role of the gravitational potential, acting on matter constituents dynamics. Moreover, equation \eqref{eqsnewpsi} shows that the strength in the matter-gravity  coupling is determined by the term $8\pi G_0\sigma$, i.e., the effective gravitational ``constant'' coupling. The field $\sigma$ is not sourced by matter neither direct acts upon matter movement as seen in \eqref{eqeuler}.


\section{Background cosmological solutions of the revisited case}

Considering the cosmological background, the function $\sigma$ is supposed to depend on time only.
Hence, the full set of equations simplify considerably to,
\begin{eqnarray}
\label{ce1}
\nabla^2 \psi + \frac{\omega}{2}\frac{\dot\sigma^2}{\sigma^2} &=& 4\pi G_0\sigma\rho,\\
\ddot\sigma - \frac{3}{2}\frac{\dot\sigma^2}{\sigma} + \frac{\dot\psi}{\psi}\dot\sigma  - \frac{\sigma}{2\omega}\frac{\nabla\psi\cdot\nabla\psi}{\psi} &=& 0,\\
\frac{\partial\rho}{\partial t} + \nabla\cdot(\rho\vec v) &=& 0,\\
\label{ce4}
\frac{\partial \vec v}{\partial t} + \vec v\cdot\nabla \vec v &=&  - \nabla\psi.
\end{eqnarray}

Since the cosmological velocity field is described by \eqref{vHr} and, again, the density (and the pressure) are functions of time only, equation (\ref{ce1}) can be written as,
\begin{eqnarray}
\psi = \frac{g(t)}{6}r^2.
\end{eqnarray}
with $g(t)$ given as previously by \eqref{g(t)}.
We then end up with two coupled equations for the scalar functions
\begin{eqnarray}
\label{fc1}
\frac{\ddot\sigma}{\sigma}  + \biggr(2H+\frac{\dot g}{g}\biggl)\frac{\dot\sigma}{\sigma} - \frac{3}{2}\frac{\dot\sigma^2}{\sigma^2} &=& \frac{g}{3\omega},\\
\label{fc2}
\dot H + H^2 &=& - \frac{g}{3}.
\end{eqnarray}

The above equation \eqref{fc2} resembles the usual Friedmann equation for a pressureless universe by setting $\sigma=1$. 

Our goal in the next steps is to find solutions for the above set of equations.

\subsection{Power-law solution}
From the conservation equation, we have the same solution $\rho = \rho_0 a^{-3}$ as in \eqref{density}. Let us now suppose the power law solution,
\begin{eqnarray}
a = a_0 t^{\alpha}, \quad
\sigma = \sigma_0 t^{\beta}.
\end{eqnarray}
The expression for $g(t)$ defined in \eqref{g(t)} becomes,
\begin{eqnarray}
g(t) = 4\pi G_0\sigma_0\frac{\rho_0}{a_0^3}t^{\beta - 3\alpha} - \frac{\omega}{2}\beta^2 t^{-2}.
\end{eqnarray}
By demanding on $g(t)$ a unique time dependence as $g(t) \propto t^{-2}$, the above solution leads to the following constraining relation,
\begin{eqnarray}
\label{sr-rel}
\beta = 3\alpha - 2.
\end{eqnarray}
Combining now (\ref{fc1}) and (\ref{fc2}), with the relation between $\alpha$ and $\beta$ above, we end up with,
\begin{eqnarray}
\biggr(1 + \frac{3}{2}\omega\biggl)\beta^2 + (1 - 15\omega)\beta  - 2 = 0,
\end{eqnarray}
with the solution,
\begin{eqnarray}
\label{root}
\beta = \frac{15\omega-1 \pm \sqrt{9 - 18\omega + 225\omega^2}}{3\omega + 2}.
\label{beta}
\end{eqnarray}

Equivalently, via \eqref{sr-rel},
\begin{eqnarray}
\label{rootr}
\alpha = \frac{21\omega + 3 \pm \sqrt{9 - 18\omega + 225\omega^2}}{3(3\omega + 2)}.
\end{eqnarray}

For $\omega = -2/3$ there is just one solution $\beta = 2/11$, and consequently $\alpha = 8/11$. For $\omega = 0$ and negative sign in eq. (\ref{root}) the scale factor is constant but the gravitational coupling decreases. 
\begin{figure}[ht]
    \centering
    \begin{minipage}[h]{0.5\columnwidth}
        \centering
        \includegraphics[width=\columnwidth]{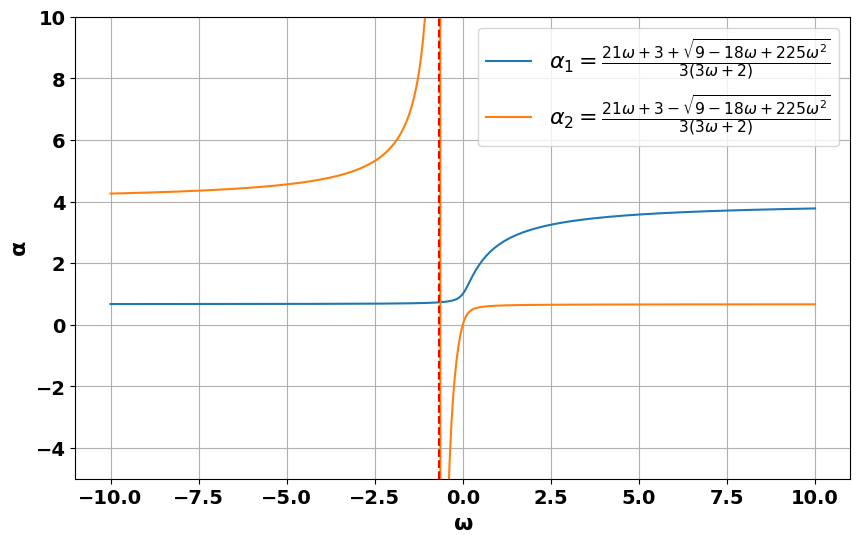}
        \label{fig:r_omega}
    \end{minipage}%
    \hfill
    \begin{minipage}[h]{0.5\columnwidth}
        \centering
        \includegraphics[width=\columnwidth]{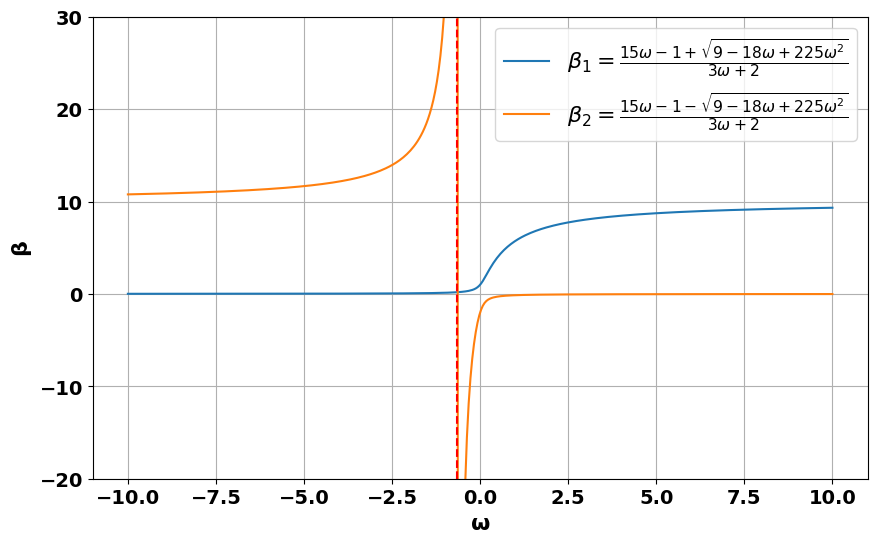}
        \label{fig:s_omega}
    \end{minipage}
    \caption{Dependence of parameters $\alpha$ and $\beta$ on the $\omega$ values [cf. eqs. (\ref{root}) and \eqref{rootr}].}
    \label{fig+}
\end{figure}

In figure \ref{fig+} we plot the roots for $\alpha$ and $\beta$ as a function of $\omega$ choosing the plus and minus sign, respectively. It can be remarked the presence of accelerated solutions for the universe. 

Using (\ref{ce1}) and (\ref{fc2}), it comes out,
\begin{eqnarray}
     - 3\alpha(\alpha - 1) + \frac{\omega}{3}\beta^2= 4\pi G_0\frac{\sigma_0\rho_0}{a_0^3}.
\end{eqnarray}
If we impose that gravity must act attractively, then $\sigma\rho >  0$. This condition leads, with the aid of (\ref{sr-rel}), to the relation
\begin{eqnarray}
    (3\omega - 2)\beta^2 - 2\beta + 4 > 0,
\end{eqnarray}
in order gravity to be attractive.
For the plus sign in eq. (\ref{root}) this condition is satisfied everywhere except in the interval $0 < \omega \lesssim 0.75$ while for the minus sign it is satisfied only for $\omega > 0$. 

\begin{figure}[ht]
    \centering
    \begin{minipage}[t]{0.5\columnwidth}
        \centering
        \includegraphics[width=\columnwidth]{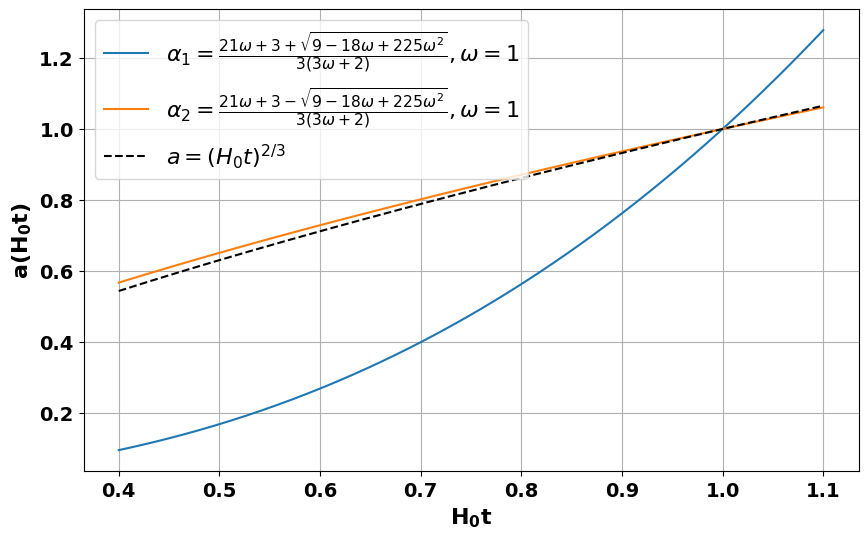}
        \label{fig:scale_factor}
    \end{minipage}%
    \hfill
    \begin{minipage}[t]{0.5\columnwidth}
        \centering
        \includegraphics[width=\columnwidth]{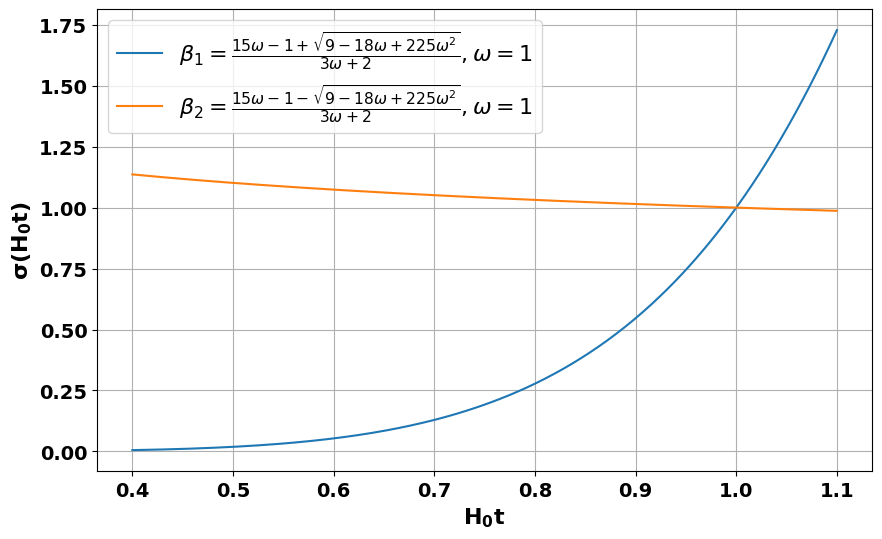}
        \label{fig:sigma_field}
    \end{minipage}
    \caption{Evolution of of the scale factor (left) and the $\sigma$ field (right) as a function of the normalized cosmic time  adopting $\omega = 1$ in \eqref{root}. This case corresponds to an attractive gravitational effect when the sign is negative and an expanding framework when the sign is positive. All functions are normalized to unity at present ($t =H_0^{-1}$).}
    \label{base-6}
\end{figure}

\begin{figure}[ht]
    \centering
    \begin{minipage}[t]{0.5\columnwidth}
        \centering
        \includegraphics[width=\columnwidth]{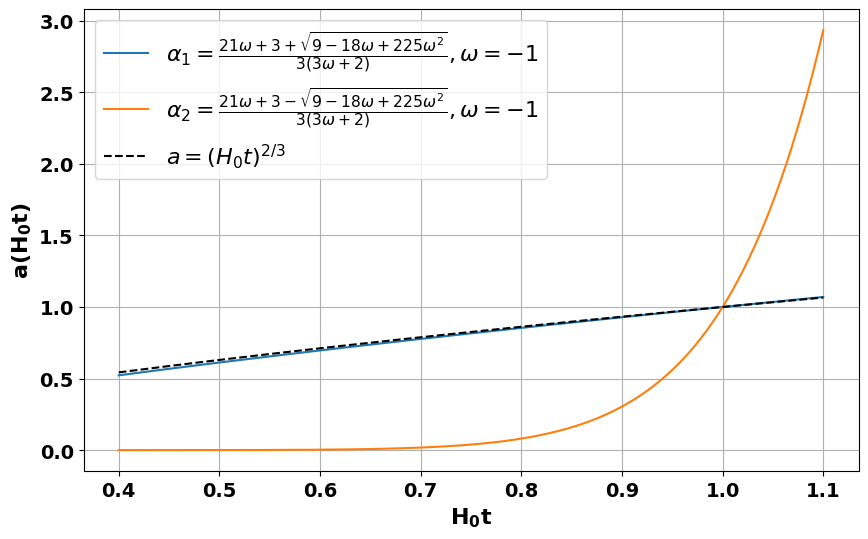}
        \label{fig:scale_factor_neg}
    \end{minipage}%
    \hfill
    \begin{minipage}[t]{0.5\columnwidth}
        \centering
        \includegraphics[width=\columnwidth]{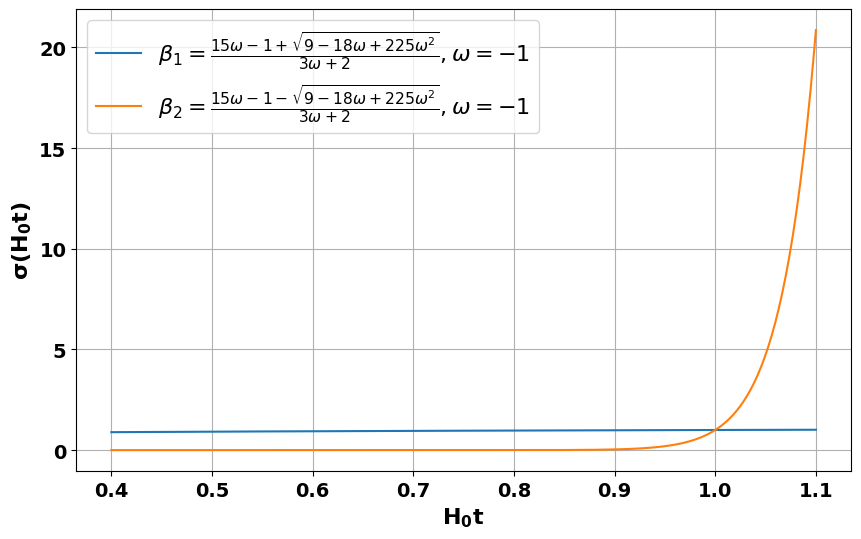}
        \label{fig:sigma_field_neg}
    \end{minipage}
    \caption{Evolution of of the scale factor (left) and the $\sigma$ field (right) as a function of the normalized cosmic time adopting $\omega = -1$ in \eqref{root}. This case corresponds to an attractive gravitational effect when the sign is positive and an expanding framework when the sign is negative. All functions are normalized to unity at present.}
    \label{base-6_neg}
\end{figure}

The possible cosmological scenarios can be summarized as follows:
\begin{itemize}
    \item For $\omega > 0$ and positive sign in eq. (\ref{root}), the  universe expands accelerated ($\alpha > 1$) with an increasing gravitational coupling ($\beta > 0$). The power law index for the scale factor is in the interval $1 \leq \alpha \leq 4$. Gravity is attractive, except in the interval $0 < \omega \lesssim 0.75$.
    \item For $\omega < 0$ and positive sign in eq. (\ref{root}), the  universe is decelerating ($\alpha < 1$) with an increasing gravitational coupling ($\beta > 1$). The power law index for the scale factor is in the interval $1 \geq \alpha \geq 2/3$. Gravity is attractive.
    \item For $\omega > 0$ and negative sign in eq. (\ref{root}), the  universe is decelerating ($\alpha < 1$) with an decreasing gravitational coupling ($\beta < 0$). The power law index for the scale factor is in the interval $0 \geq \alpha \geq 2/3$. Gravity is attractive.
    \item For $\omega < 0$ and negative sign in eq. (\ref{root}), the  universe is always accelerating ($\alpha > 1$ or $\alpha < 0$) with an increasing gravitational coupling ($- \infty < \omega < - 2/3$) or decreasing gravitational coupling $- 2/3 \geq \omega > 0$. Gravity is repulsive.
\end{itemize}
The evolution of the scale factor and the $\sigma$ field are depicted in figures \ref{base-6} ($\omega=1$) and \ref{base-6_neg} ($\omega=-1$). In both left panels of these figures the dashed line corresponds to the evolution of Einstein-de Sitter universe ($a\propto t^{2/3}$). 

\subsection{Mimicking the $\Lambda$CDM model}

The power law solution proposed in the last section is quite limited since it can not provide a smooth transition from different cosmological eras as desired by the current available observational data. This is however also true even in General Relativity i.e., a single fluid power law solution can not represent a full cosmological model since it is not able to provide a smooth transition between the different cosmological eras.

In order to find a suitable model which is compatible with current available observational data on the cosmological background evolution a good strategy is to remain close to the standard cosmological model given by the $\Lambda$CDM expansion. We therefore seek now a solution based on the $\Lambda$CDM model which obviously is compatible with observational data.  Thus, we look for the $\sigma(t)$ function that provides the same evolution as the standard flat $\Lambda$CDM model. 


Starting from the second Friedmann equation,
\begin{equation}
\label{F2}
    \frac{\ddot{a}}{a} = -\frac{4\pi G}{3}(\rho + 3p)
\end{equation}
we obtain, by combining \eqref{F2} with \eqref{g(t)} through \eqref{fc2},
\begin{equation}
\label{rhom}
    4\pi G \sigma\rho_m - \frac{\omega}{2}\frac{\dot{\sigma}^2}{\sigma^2} = 4\pi G(\rho + 3p)
\end{equation}

Assuming $\rho = \rho_m$ on the left-hand side of the equation above---specifically in equation \eqref{g(t)}---is based on the premise that the variable $G$ model proposed here is supported by a universe dominated solely by matter. In this case, $\sigma$ would be responsible for the expansion in the same way that $\rho_\Lambda$ is in $\Lambda$CDM. Thus, considering that we expect this model to reproduce the same behavior as $\Lambda$CDM, we will adopt for the density $\rho$ and pressure $p$ terms described by \eqref{F2} (and also present on the right-hand side of \eqref{rhom}): $p=p_R+p_m+p_{\Lambda }=\rho_r/3 - \rho_\Lambda$. After some algebraic manipulation and solving for $\dot{\sigma}(t)$, we obtain
\begin{equation}
\label{sigmadot2}
\dot\sigma = \pm H_0\sigma\sqrt{\frac{3}{\omega}\left(2\Omega_\Lambda + \sigma\frac{\Omega_{m0}}{a^3} - \frac{\Omega_{m0}}{a^3}\right)}.
\end{equation}

\begin{figure}[ht]
    \centering
    \begin{minipage}[t]{0.47\columnwidth}
        \centering
        \includegraphics[width=\columnwidth]{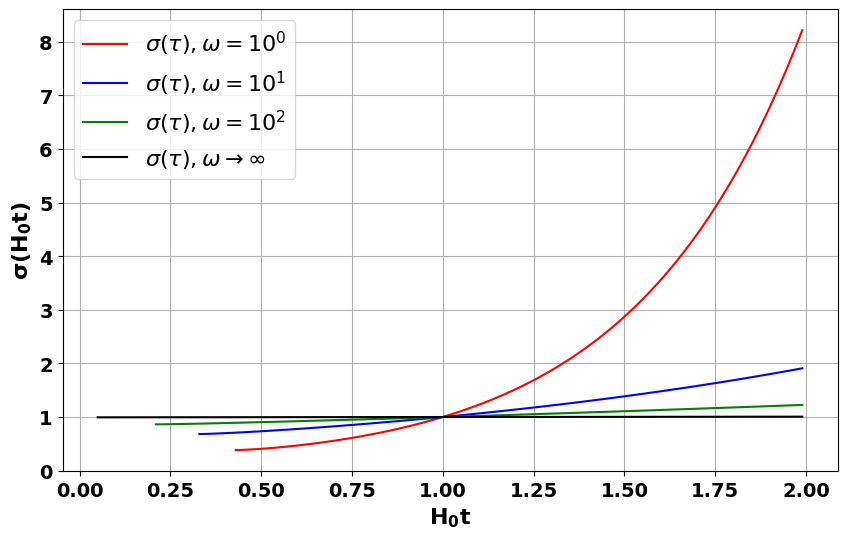}
        \label{fig:sigmaHt}
    \end{minipage}%
    \hfill
    \begin{minipage}[t]{0.5\columnwidth}
        \centering
        \includegraphics[width=\columnwidth]{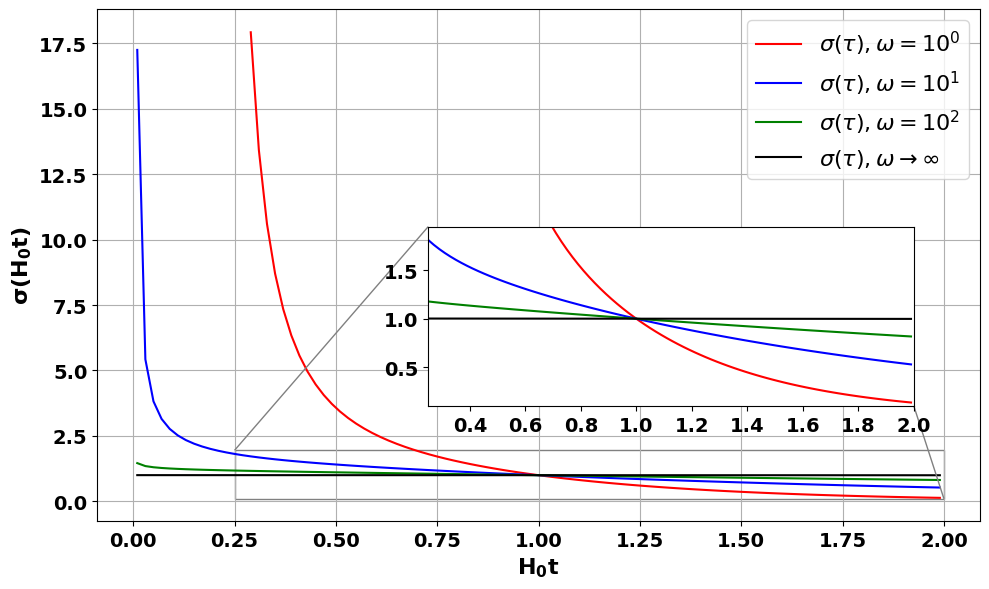}
        \label{fig:sigmaa}
    \end{minipage}
    \caption{Evolution of the field $\sigma$ as a function of the normalized cosmic time. The left (right) panel adopts a positive (negative) sign in the right hand side of \eqref{sigmadot2}.}
    \label{figsigma}
\end{figure}

The above equation can not be solved analytically. However, since this is a first order differential equation, its integration constant can be adjusted to match the available limits on $\dot{G}/G$ (interpreting this quantity as proportional to $\dot{\sigma}/\sigma$) as given my the MICROSCOPE mission \cite{microscope}. Using the results from Ref. \cite{microscope}, and defining the dimensionless logarithmic variation of $G$ divided by the Hubble parameter today, we arrive at $\dot{G}/H_0 G < 10^{-5}$. Therefore, the gravitational coupling today is nearly constant. To make some estimates, let us fix $\dot{\sigma}/\sigma |_{today} \sim 0$ today. Thus, using this approximation we solve numerically \eqref{sigmadot2}. After numerical integration of \eqref{sigmadot2}, the behavior of the scalar field $\sigma$ as a function of $H_0t$ is shown in Fig. \ref{figsigma} for different values of $\omega$. The limit $\omega \rightarrow \infty$, the green line in both panels of this figure, corresponds to a constant $\sigma$ behavior as in GR.

\begin{figure}[ht]
    \centering
    \begin{minipage}[t]{0.49\columnwidth}
        \centering
        \includegraphics[width=\columnwidth]{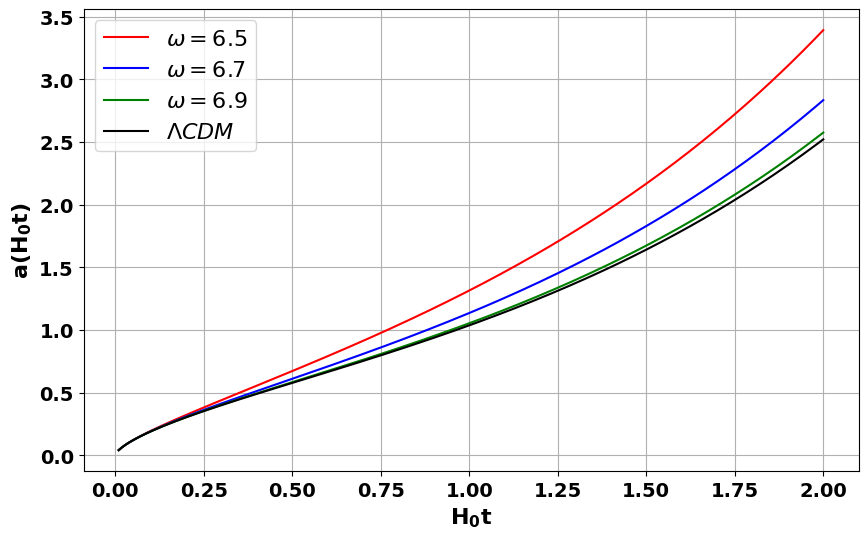}
    \end{minipage}%
    \hfill
    \begin{minipage}[t]{0.5\columnwidth}
        \centering
        \includegraphics[width=\columnwidth]{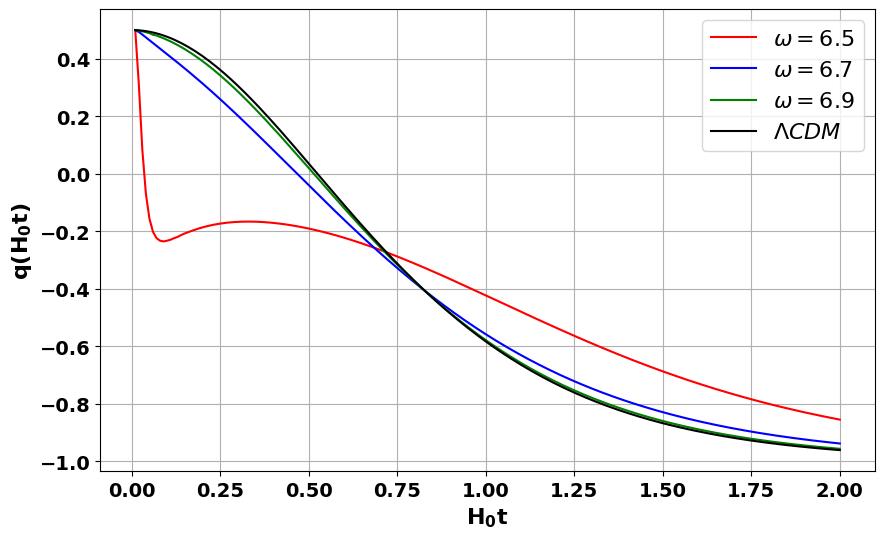}
    \end{minipage}
    \caption{Evolution of the scale factor (left) and the deceleration parameter (right) as a function of the normalized cosmic time. In both panels we adopt a negative sign in the right hand side of \eqref{sigmadot2}.}
    \label{fig:aandq}
\end{figure}

Having now the evolution of the $\sigma$ field we can solve \eqref{fc2} to obtain the scale factor. In Figure \ref{fig:aandq} we show the scale factor and the deceleration parameter as a function of the normalized cosmic time for different values of the $\omega$ parameter adopting the negative sign in the right hand side of \eqref{sigmadot2} which is the most plausible case analysed. We have verified that positive sign in the right hand side of \eqref{sigmadot2} leads to meaningless solutions. This is the most constraining test so far. In order to remain close to the $\Lambda$CDM model i.e., providing a smooth transition from the matter behavior, where $q=+0.5$, to the asymptotic future de Sitter expansion, where ($q=-1$), the free model parameter $\omega$ should obey $\omega \gtrsim 6.7$.

\section{Perturbations}
In the standard Newtonian theory, the evolution of pressureless matter density perturbations is given by the equation \cite{Weinberg:1972kfs}
\begin{eqnarray}
    \ddot\delta + 2 H\dot\delta - 4\pi G_0\rho\delta = 0,
\end{eqnarray}
where $\delta$ is the density contrast defined by the fluctuation of on the density divided by the background density:
\begin{eqnarray}
    \delta = \frac{\delta\rho}{\rho}.
\end{eqnarray}
The scale factor behaves as $a \propto t^{2/3}$ the same behaviour as the matter dominated cosmological model using general relativity. The solution for the density contrast is therefore given by,
\begin{eqnarray}
    \delta \propto t^{2/3},
\end{eqnarray}
which coincides also with the general relativity result \cite{Weinberg:1972kfs}.

For the sake of comparison, using the McVittie approach to a $G$ variable cosmological model \cite{McVittie:1978}, the scale factor is given by $a \propto t^{1/3}$, and the density contrast was computed in Ref. \cite{baptista1984gravitational} \nocite{Baptista}, leading to,
\begin{eqnarray}
    \delta \propto t^p, \quad p = 1.
\end{eqnarray}

We now return to the Newtonian theory with variable $G$ introduced earlier. To facilitate a perturbative analysis around the previously derived solutions, it is more convenient to redefine the field $\sigma$ as follows
\begin{eqnarray}
\sigma = e^\phi.
\end{eqnarray}
With this redefinition, the equations become,
\begin{eqnarray}
\nabla^2 \psi - \nabla\psi\cdot\nabla\phi + \frac{\omega}{2}\dot\phi^2 &=& 4\pi G_0e^\phi\rho,\\[1ex]
\ddot\phi - \frac{1}{2}\dot\phi^2 + \frac{\dot\psi}{\psi}\dot\phi + c^4 \frac{e^{2\phi}}{\psi} \biggr(- \nabla^2\phi
- \frac{\nabla\phi\cdot\nabla\phi}{2}\biggl) - \frac{1}{2\omega}\frac{\nabla\psi\cdot\nabla\psi}{\psi} &=& 0,\\[1ex]
\frac{\partial\rho}{\partial t} + \nabla\cdot(\rho\vec v) &=& 0,\\[1ex]
\frac{\partial \vec v}{\partial t} + \vec v\cdot\nabla \vec v &=& - \frac{\nabla p}{\rho} - \nabla\psi,
\end{eqnarray}
and, after performing a linear perturbation and using background equations to simplify, one obtains
\begin{eqnarray}
\label{pe1a}
\nabla^2 \delta\psi - \nabla\psi\cdot\nabla\delta\phi + \omega\dot\phi\delta\dot\phi &=& \biggr(g + \frac{\omega}{2}\dot\phi^2\biggl)\biggr\{\delta\phi + \frac{\delta\rho}{\rho}\biggl\},\\[1ex]
\label{pe2a}
\delta\ddot\phi - \dot\phi\delta\dot\phi + \biggr(\frac{\delta\dot\psi}{\psi} - \frac{\dot\psi\delta\psi}{\psi^2}\biggl)\dot\phi + \frac{\dot\psi}{\psi}\delta\dot\phi  - c^4\frac{e^{2\phi}}{\psi}\nabla^2 \delta\phi &=& \frac{1}{\omega}\frac{\nabla\psi\cdot\nabla\delta\psi}{\psi} - \frac{1}{2\omega}\frac{\nabla\psi\cdot\nabla\psi}{\psi}\frac{\delta\psi}{\psi},\\[1ex]
\label{pe3a}
\dot\delta + \nabla\cdot\delta \vec v &=& 0,\\[1ex]
\label{pe4a}
\delta\dot{\vec v} + H\delta \vec v &=& - c_s^2\nabla\delta  - \nabla\delta\psi.
\end{eqnarray}
In the above expressions we have defined the density contrast and the sound velocity, respectively
\begin{eqnarray}
\delta = \frac{\delta\rho}{\rho}, \quad c_s^2 = \frac{\partial p}{\partial \rho},
\end{eqnarray}
and an upper dot means total time derivative, namely
\begin{eqnarray}
\dot f = \frac{\partial f}{\partial t} + \vec v\cdot\nabla f.
\end{eqnarray}
Combining the last two equations, the system of coupled perturbed equations can be simplified further.
Using also the background equations, the perturbed equations become,
\begin{eqnarray}
\label{pe1b}
\nabla^2 \delta\psi - \nabla\psi\cdot\nabla\delta\phi + \omega\dot\phi\delta\dot\phi &=& \biggr(g + \frac{\omega}{2}\dot\phi^2\biggl)\biggr\{\delta\phi + \frac{\delta\rho}{\rho}\biggl\},\\[1ex]
\label{pe2b}
\delta\ddot\phi - \dot\phi\delta\dot\phi + \biggr(\frac{\delta\dot\psi}{\psi} - \frac{\dot\psi\delta\psi}{\psi^2}\biggl)\dot\phi + \frac{\dot\psi}{\psi}\delta\dot\phi  - c^4\frac{e^{2\phi}}{\psi}\nabla^2 \delta\phi &=& \frac{1}{\omega}\frac{\nabla\psi\cdot\nabla\delta\psi}{\psi} - \frac{1}{2\omega}\frac{\nabla\psi\cdot\nabla\psi}{\psi}\frac{\delta\psi}{\psi},\\[1ex]
\label{pe3b}
\ddot\delta + 2H\dot\delta - c_s^2\nabla^2\delta &=& \nabla^ 2\delta\psi.
\end{eqnarray}


The system of equations (\ref{pe1b}-\ref{pe3b}) does not admit a Fourier decomposition. This can be seen from the first two terms of (\ref{pe1b}). The first one implies a function only of time, while the second one will carry a term $\vec r\cdot \vec k$, which also depends on the position. The right-hand-side of (\ref{pe2b}) is similarly problematic in this sense. In fact, a Fourier decomposition in linear perturbative analysis is not always possible, see for example Ref. \cite{bronnikov2021simplest}.

In order to circumvent this issue, we will suppose that the perturbations have a behavior similar of the background:
\begin{eqnarray}
\delta\psi &=& F(t) \frac{r^2}{6},\\
\delta\phi &=& \delta\phi(t),\\
\delta\vec v &=& V(t)\vec r,\\
\delta &=& \delta(t).
\end{eqnarray}

We obtain, after some manipulations, two coupled differential equations, namely,
\begin{eqnarray}
\label{p1}
(3 - 2A)\delta\ddot\phi + \biggr\{\frac{\dot g}{g} + 2H- \frac{2}{3}\dot\phi + A\dot\phi - \omega\dot\phi\frac{d}{dt}\biggr(\frac{\dot\phi}{g}\biggl) \biggl\}\delta\dot\phi 
+ \biggr\{\dot A \dot\phi - \frac{g}{3\omega}A\biggl\}\delta\phi &=&
- A\dot\phi\dot\delta + \biggr\{- \dot A\dot\phi + \frac{A}{3\omega}g\biggl\}\delta,\\[1ex]
\label{p2}
\ddot\delta + 2H\dot\delta - gA\delta &=& gA\delta\phi - \omega\dot\phi\delta\dot\phi.
\end{eqnarray}
In these expressions we have defined,
\begin{eqnarray}
A = 1 + \frac{\omega}{2}\frac{\dot\phi^2}{g}.
\end{eqnarray}
It is worth to mention that for the power law background solutions found above definition  becomes a constant.

\begin{figure}[h]
\centering
\includegraphics[width=0.48\linewidth]{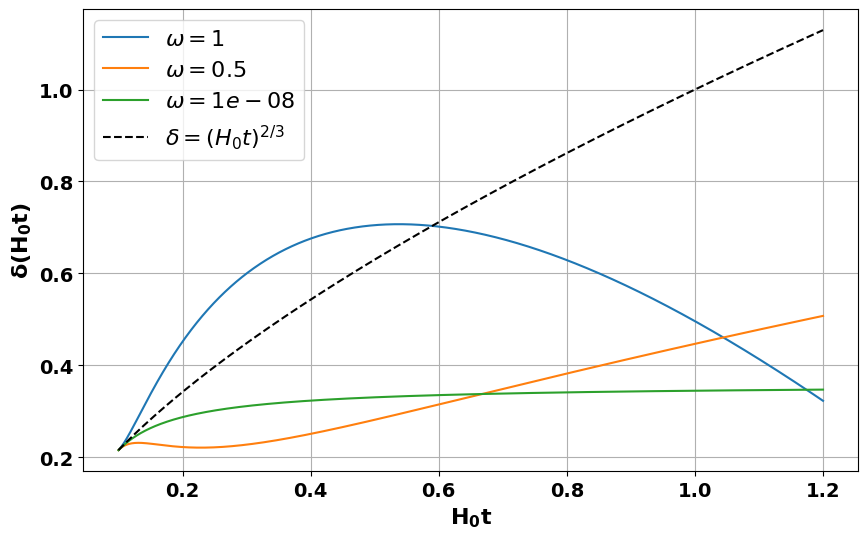}\qquad
\includegraphics[width=0.48\linewidth]{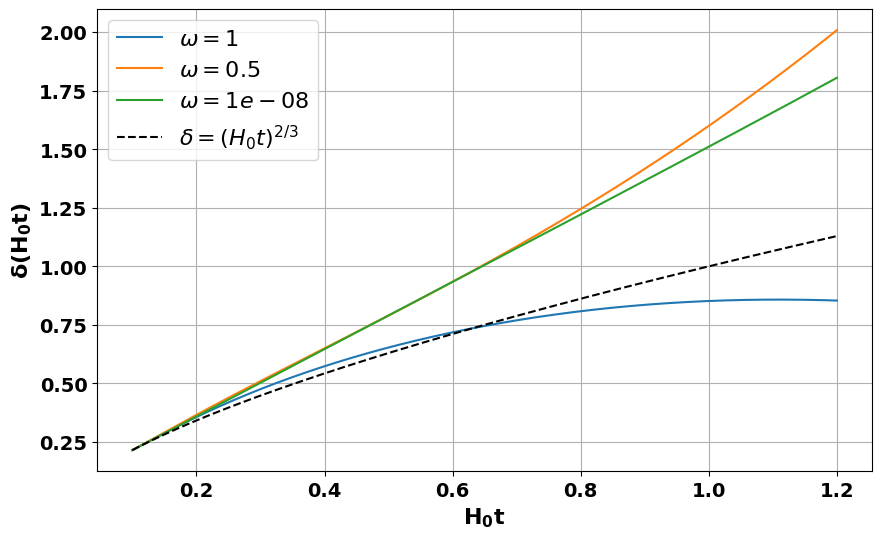}
\caption{Evolution of $\delta(t)$ in the variable Newtonian $G$ theory (continuous curves) for different values of $\omega$. On the left for the plus sign in \eqref{beta} and on the right for the negative sign in \eqref{beta}. The usual Newtonian theory result is represented by the dashed line.} 
\label{fig-omega>0}
\end{figure}

Using the power law solutions for the  background, the equations become,
\begin{eqnarray}
\label{p1a}
a_1\delta\ddot\phi + a_2\frac{\delta\dot\phi}{t}
+ a_3\frac{\delta\phi}{t^2} &=&
b_1\frac{\dot\delta}{t} + b_2\frac{\delta}{t^2},\\[1ex]
\label{p2a}
\ddot\delta + b_3\dot\delta + b_ 4\frac{\delta}{t^2} = a_4\frac{\delta\dot\phi}{t} + a_5\frac{\delta\phi}{t^ 2},
\end{eqnarray}
where the $a_i, b_i$ are constants coefficients depending on $\alpha, \beta$ and $\omega$.
Remark that $a_3 = - b_2$ and $a_5 = - b_4$.
These coefficients read:
\begin{eqnarray}
    a_1 &=& 1 + 3\omega\frac{\beta^2}{(\beta + 2)(\beta - 1)},\\
    a_2 &=& 3\beta - \frac{2}{3} - \frac{3}{2}\omega\frac{\beta^2(\beta - 2)}{(\beta + 2)(\beta - 1)}\\
    a_3 &=& - \frac{\beta^2}{6} + \frac{(\beta + 2)(\beta -1)}{9\omega},\\
    a_4 &=& - \omega \beta,\\
    a_5 &=& - 3\omega a_3,\\
    b_1 &=& - \beta + \frac{3}{2}\omega\frac{\beta^3}{(\beta + 2)(\beta - 1)},\\
    b_2 &=& - a_3,\\
    b_3 &=& \frac{2}{3}(\beta + 2),\\
    b_4 &=& - a_5.
\end{eqnarray}

\begin{figure}[h]
    \centering
    \begin{minipage}[t]{0.48\columnwidth}
        \centering
        \includegraphics[width=\columnwidth]{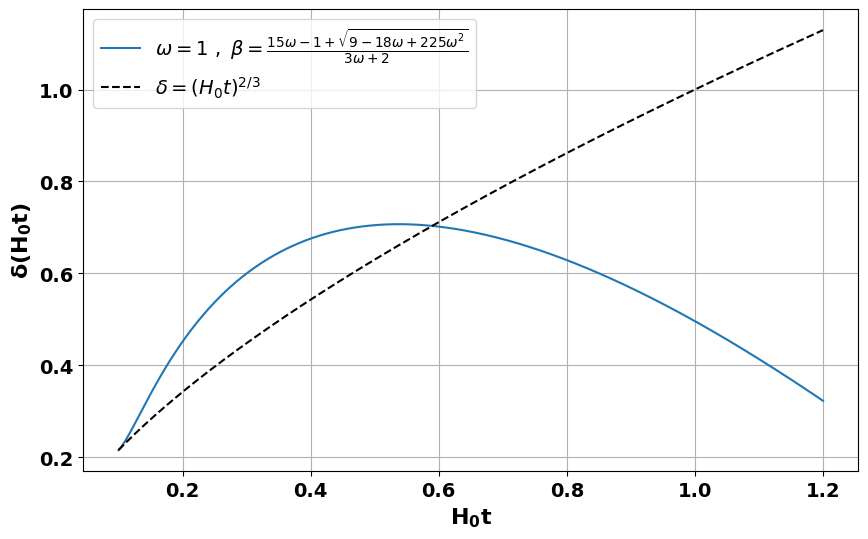}
        \label{fig:delta1+1}
    \end{minipage}%
    \hfill
    \begin{minipage}[t]{0.48\columnwidth}
        \centering
        \includegraphics[width=\columnwidth]{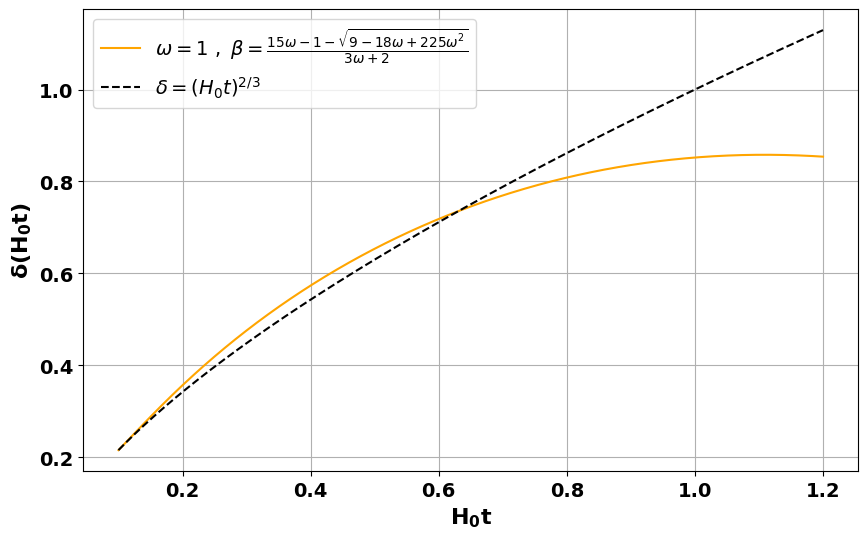}
        \label{fig:delta1-1}
    \end{minipage}
    \vfill
    \begin{minipage}[t]{0.48\columnwidth}
        \centering
        \includegraphics[width=\columnwidth]{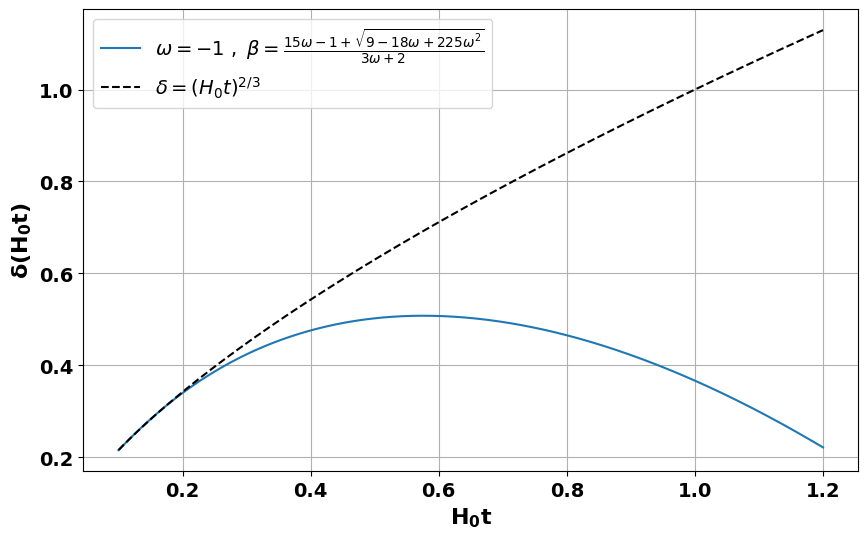}
        \label{fig:delta-1+1}
    \end{minipage}%
    \hfill
    \begin{minipage}[t]{0.48\columnwidth}
        \centering
        \includegraphics[width=\columnwidth]{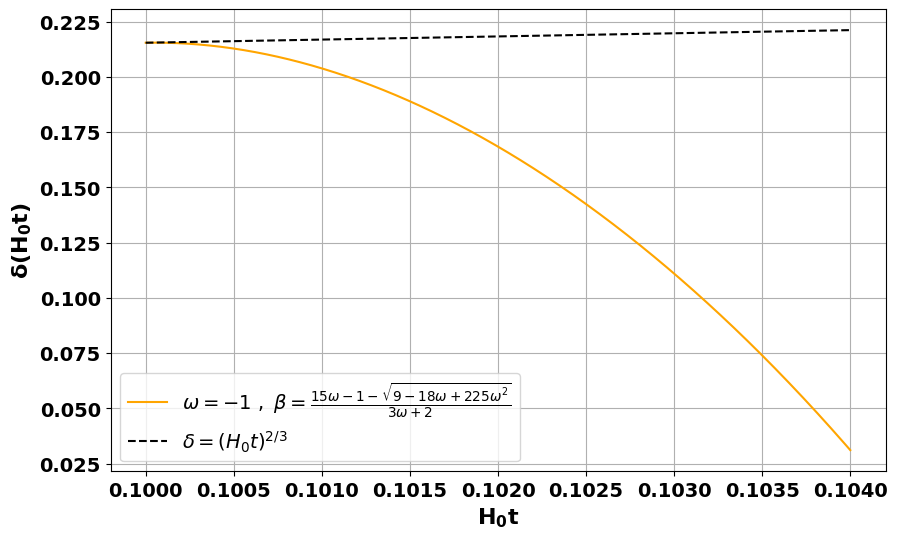}
        \label{fig:delta-1-1}
    \end{minipage}
    \caption{Evolution of $\delta(t)$ in the variable Newtonian $G$ theory (continuous curves) for $\omega = 1$ and $\omega = -1$. The usual Newtonian theory result is represented by the dashed line.}
    \label{fig-delta1}
\end{figure}

Examples of the evolution of the density contrast function, compared with the usual Newtonian case ($\delta \propto t^{2/3}$), are displayed in figures \ref{fig-omega>0} and \ref{fig-delta1}.

It is possible to obtain exact solutions for the coupled equations
(\ref{p1a}) and (\ref{p2a}). These equations constitute a system of coupled Euler-type equations. Power law solutions can be obtained under the form,
\begin{eqnarray}\label{deltap}
    \delta = \delta_0 t^p, \quad \phi = \phi_0 t^p.
\end{eqnarray}
In principle a fourth order equation for $p$ is obtained, but it reduces to third order algebraic equation in view of the relations for $a_3, a_5, b_2$ and $b_4$. We will refer to $p_i$ with $\{i=1,2,3\}$ as the possible three solutions for $p$ in \eqref{deltap}.  A variety of solutions is given in the table \ref{t1} in terms of $\omega$ and the sign chosen in eq. (\ref{root}).
\begin{table}[h]
\centering
\renewcommand{\arraystretch}{1.25}
\caption{Density contrast exponents for specific choices of $\omega$ and the sign of eq. \eqref{root}.}
\label{t1}
\begin{tabular}{|c|c|c|c|}\hline\hline
$\omega$/Sign in eq. (\ref{root})&$p_1$&$p_2$&$p_3$  \\ \hline\hline
$\omega = -1 $/Sign $-$& $-29.99$&$7.02$&$41.44$\\ \hline
$\omega = -1 $/Sign $+$& $-1.01$&$0.96$&$1.13$\\ \hline
$\omega = +1 $/Sign $-$& $-0.99$&$0.56$&$2.05$\\ \hline
$\omega = +1 $/Sign $+$& $-4.54$&$0.68+0.30i$&$0.68-0.30i$\\ \hline
$\omega = -0.1 $/Sign $-$& $-1.27$&$2.32$&$14.21$\\ \hline
$\omega = -0.1 $/Sign $+$& $-1.03$&$0.59+0.85i$&$0.59 - 0.85i$\\ \hline
$\omega = +0.1 $/Sign $-$& $-0.94$&$0.55$&$4.43$\\ \hline
$\omega = +0.1 $/Sign $+$& $-0.85$&$-0.12+1.04i$&$-0.12-1.04i$\\ \hline
\end{tabular}
\end{table}

We have found a a variety of possible solutions, including, in general, a faster grow of perturbations in both decelerated and and accelerated backgrounds. Concerning the latter, the most surprising result in the table \ref{t1} corresponds to the  first line, with the choice of minus sign in eq. (\ref{root}) and $\omega < 0$. In this case, the scale factor exhibits accelerated expansion and the perturbations grow faster than in the standard case with $G$ constant. If we have chosen $\omega = -0.3$, for example, $\alpha = 1.6$ (accelerated expansion) and $p \gtrsim 1$, implying an enhanced growth of perturbations. Remark also that, in this scenario, the gravitational coupling is also growing.

\section{Conclusions}

The Newtonian theory with variable G developed here differs from the previous formulation \cite{Escorcio:2023gnu} in that the field $\sigma$, representing the dynamic gravitational coupling, is now associated with the gravitational potential rather than the matter sector. This approach is more closely aligned with the original Brans–Dicke theory. Nevertheless, both formulations can be related through a field redefinition, analogous to a conformal transformation in the relativistic Brans–Dicke framework, in which the non-minimal coupling between the scalar field and gravity can be removed at the expense of introducing a non-minimal coupling within the matter sector.

Considering the homogeneous and isotropic cosmological background expansion, both formulations lead to similar qualitative results. However, the formulation presented in Ref. \cite{Escorcio:2023gnu} does not admit power law solutions, differently from what happens in the new approach presented in the present work. At perturbative level, the main technical difficulty in the new approach comes from the impossibility of performing a Fourier decomposition after linearization of the dynamical equations: this is due to the direct coupling of the gravitational potential and the scalar field $\sigma$. On the other hand, the new formulation discussed in the present text preserves the Euler equation, which is connected to the geodesic equation in a possible relativistic formulation, again close to the original spirit of the Brans-Dicke theory which has a non-minimal coupling between scalar field and gravity but preserves the geodesic motion. We have also found that there is a qualitative similarity with the $\Lambda$CDM model for a lower bound $\omega \gtrsim 6.7$ on the coupling parameter. Of course, a more detailed statistical analysis is demanded in order to place more accurate bounds on $\omega$. We let this task for a future work. 

Is there a relativistic version for the theory developed here or in the Refs. \cite{Escorcio:2023gnu}? All extensions of General Relativity theory using scalar field leads to the Newtonian theory in the weak field limit, even in the Horndesky class of theories \cite{Kobayashi_2019}.
A construction of a relativistic version of the Newtonian theory developed here requires new type of coupling not included in the most common formulation of scalar-tensor theories.

The key novelty of this Newtonian theory with variable $G$ is its ability to produce accelerated expansion without the need for dark energy, with the cosmic acceleration driven solely by the evolving gravitational coupling. Additionally, certain configurations allow for a significant amplification of density perturbations (cf. \cite{10.1093/mnras/183.4.749,Baptista}). This might represent a viable mechanism to generate huge density clustering amplifications early on in the first stage of the structure formation process. At this stage, we can only speculate that this mechanism could shed some light in the issue of the high-z galaxies found by the JWST, but future investigation is demanded. See also Refs. \cite{10.1093/mnras/stad2448,vanPutten:2023ths} for a discussion on this topic.

We highlight that the primary goal of this study is to formulate a Newtonian counterpart to covariant scalar-tensor theories, notably the Brans-Dicke theory. We show that this Newtonian framework not only exists but also enables the investigation of several features of relativistic cosmology through the lens of a significantly simpler theory, closely aligned with classical Newtonian gravity.

\bigskip

{\bf Acknowledgements:} We thank the Brazilian research funding agencies CNPq, FAPEMIG and FAPES for partial financial support. 

{\bf Data Availability Statement:} The data that support the findings of this study are available from the corresponding author, upon reasonable request

\bibliography{Refs}
\end{document}